\documentclass[aps,prl,preprintnumbers,amsmath,amssymb,twocolumn,superscriptaddress,showpacs,floatfix]{revtex4-1}
\usepackage{amsmath,amssymb,amsfonts}
\usepackage{float}
\usepackage{hyperref}
\usepackage{graphicx}
\usepackage{xcolor}
\usepackage{tensor}
\usepackage{hyperref}
\usepackage[normalem]{ulem}
\usepackage{todonotes}
\begin{document}
\title{Suppression of thermal vorticity as an indicator of QCD critical point}
\author{Sushant K. Singh}
\email[Correspondence email address: ]{sushant7557@gmail.com}
\affiliation{Variable Energy Cyclotron Centre, 1/AF, Bidhan Nagar , Kolkata, India}
\affiliation{HBNI, Training School Complex, Anushakti Nagar, Mumbai 400085, India}
\author{Jan-e Alam}
\affiliation{Variable Energy Cyclotron Centre, 1/AF, Bidhan Nagar , Kolkata, India}
\affiliation{HBNI, Training School Complex, Anushakti Nagar, Mumbai 400085, India}
\date{\today}
\begin{abstract}
We study the impact of the QCD critical point (CP) on the spin polarization of $\Lambda$-hyperon generated by the thermal vorticity in viscous quark gluon plasma (QGP). 
The equations of the relativistic causal viscous hydrodynamics have been solved numerically 
in (3+1) dimensions to evaluate the thermal vorticity. The effects of the CP have been incorporated 
through the equation of state (EoS) and the scaling behavior of the  transport coefficients. 
A significant reduction in the global polarization has been found as the CP is approached. 
A drastic change induced by the CP in the rapidity dependence of the spin polarization is observed 
which can be used as a signature of the CP.
\end{abstract}
\maketitle

The results from lattice quantum chromodynamics (QCD)  and effective 
field theoretical models at non-zero temperature ($T$)  
and baryon chemical potential ($\mu$) reveal a rich and complex phase 
diagram~\cite{bazavov}. 
While at high $T$ and low $\mu$ $(\rightarrow 0)$  
the quark-hadron transition is a crossover, at low $T$ and high $\mu$  
the transition is of first order in nature. Therefore, it is expected that between the 
crossover and the first order transition there exists a point in the 
$\mu-T$ plane called the Critical End Point or simply the Critical Point (CP) 
where the first order transition ends and crossover begins~\cite{Fodor:2004nz}.  
The location of CP is not yet known from first principles~\cite{lqcd}, however, phenomenological studies 
indicate its existence~\cite{stephanov}. 
The search for the CP in 
the system formed in collisions of 
nuclei at relativistic energies is one of the outstanding problem.
At present the general consensus is that the collisions of nuclei at  
the top Relativistic  Heavy Ion Collider (RHIC) and Large Hadron Collider (LHC)
energies produce QGP with small $\mu$ and high $T$ which reverts to hadronic 
phase via a crossover.  The ongoing Beam Energy Scan - II program at RHIC and
the upcoming Compressed Baryonic Matter experiment at 
Facility for Anti-proton and Ion Research (FAIR) and 
Nuclotron based Ion Collider fAcility (NICA) are
planned to  create system of quarks and gluons 
with different $\mu$ and $T$ by varying the collision energies to 
explore the region close to the CP. 
The non-monotonic variation of the 
fluctuations in multiplicity with collision energy in the center of 
mass frame ($\sqrt{s_{NN}}$)~\cite{nxu}, the change 
of sign of the fourth cumulant of order parameter with the variations of rapidity (rapidity scan) 
~\cite{yyin1}
and beam energy~\cite{stephanovprl1}, the appearance of negative sign 
in the kurtosis of the order parameter fluctuation near the CP~\cite{stephanovprl2} are
some of the proposed signals of the CP (see~\cite{yyin} for a review and references therein). 

Relativistic hydrodynamics has been used to describe the 
space-time evolution of QGP  and  explain various experimental data 
quite successfully. One such crucial observable is the polarization of 
$\Lambda$-hyperon~\cite{LambdaStar} generated by the thermal vorticity
during the hydrodynamic evolution of the QGP. 
Invigorating theoretical activities have been witnessed 
(see~\cite{becattini2020} for a review) to understand
various aspects of the polarization within the scope
of hydrodynamics and transport models
after the experimental measurement of the
global polarization of the $\Lambda$ hyperon~\cite{LambdaStar}.

The magnetization of uncharged objects induced by mechanical
rotation, called Barnett effect~\cite{barnett} and its inverse, 
that is, the rotation generated by varying magnetization, called 
the Einstein-de Haas
effect~\cite{deHaas}
originate due to the conversion between spin ($S$) and 
orbital angular momentum ($L$) via spin-orbit 
coupling constrained by the conservation of total angular momentum $\vec{J}\ (=\vec{L}+\vec{S})$.  
Similar kind of coupling between $L$ and $S$ 
in the system formed in relativistic heavy ion collision
results into the spin polarization of particles.   
The  initial orbital angular momentum (OAM) imparted by the spectators in non-central
heavy ion collisions makes the fireball of QGP to rotate 
and polarize the quarks~\cite{ztliang}. This rotation 
may then appear as local vorticities in the fireball, 
the exact mechanism for which is not yet fully understood. Vorticity is a measure of the 
local spinning of fluid elements. 
The coupling of fluid vorticity and quantum mechanical spin has been experimentally 
demonstrated for the first time in Ref.~\cite{spinhydro}. Such an information is reflected in the 
spin polarization of final state hadrons. However, the vorticity can be generated by the viscous stresses 
of the system even in the absence of an initial OAM. Hence, spin polarization of hadrons has two 
contributions: one coming through OAM and another generated through viscosities of the system. 
The first contribution depends on the details of mechanism of transfer of initial OAM to vorticity and is 
sensitive to the initial condition. The second contribution depends on the transport properties of the system 
and will be sensitive to the EoS. 
In this letter, we focus on the second contribution and discuss about the first one in the supplemental material~\cite{supplement}. 
The goal here is to understand the CP induced change in the
vorticity and its consequences  on the spin polarization
of $\Lambda$ hyperon. In other words if
$\varpi_{\mu\nu}^{\text{CP}}$  ($\varpi_{\mu\nu}$)
is the vorticity in the presence (absence) of CP
then  what is the value of $\Delta\varpi_{\mu\nu}=
(\varpi_{\mu\nu}^{\text{CP}}-\varpi_{\mu\nu}$) and
the corresponding change on the spin polarization of $\Lambda$-hyperon.
We show that as the CP is approached the local vorticity 
and hence the polarization effect is suppressed. 

The presence of CP in the EoS affects 
the expansion of the system due to suppression of the sound wave and divergence of some of the transport coefficients~\cite{hasan}. 
This will affect the evolution of local vorticity and hence the $\Lambda$-polarization through vorticity-spin coupling. 
Apart from polarization, the effect of CP on the separation of baryon and anti-baryon 
due to chiral vortical effect is another interesting facet~\cite{cve}.  

Here we use natural unit $c=\hbar=k_B=1$ where $c$ is the speed of light in vacuum, $h=2\pi\hbar$ 
is the Planck's constant and $k_B$ is the Boltzmann's constant. The signature metric 
for flat space time is taken as $g_{\mu\nu}=\text{diag}(1,-1,-1,-1)$.

We numerically  solve (3+1)-dimensional 
relativistic viscous 
causal hydrodynamics using the algorithm detailed in Ref.~\cite{karpenko2014}. The code contains
the effect of CP through the EoS and the scaling behavior of the transport
coefficients. The  initial condition and the EoS models that we use 
to solve hydrodynamic equations have been extensively tested by reproducing the results 
available in Refs.~\cite{chunshen2020} and ~\cite{parotto2020} respectively. 
The CORNELIUS code~\cite{Cornelius} has been used to find the constant energy-density hyper surface.
Our numerical results in the absence of CP have been contrasted with the known analytical 
results of Ref.~\cite{gubser2010} and with numerical results from other publicly available codes: 
AZHYDRO~\cite{azhydro}, MUSIC~\cite{music} and 
vHLLE~\cite{karpenko2014}. The reliability of our code can be 
further appreciated by contrasting its output with the transverse momentum, rapidity and azimuthal
angle dependence of various experimental observables 
(see the supplemental material~\cite{supplement} for details).

The relativistic hydrodynamic equations that we solve are: 
\begin{align}
\partial _{\mu}T^{\mu \nu}=0, \nonumber \\
\label{charge_conservation}  \partial _{\mu}N^{\mu}=0,
\end{align}
where $T^{\mu \nu}$ is the energy-momentum tensor and $N^{\mu}$ is the net-baryon number current. 
Here we work in the Landau frame of reference where the $T^{\mu \nu}$ and $N^\mu$ are given by 
\begin{align}
T^{\mu \nu} &= \varepsilon \,u^\mu\,u^\nu - (p+\Pi)\Delta^{\mu \nu} +\pi^{\mu \nu},\\
N^\mu &= n_B\, u^\mu +V^\mu,
\end{align}
where $\Pi$ is the bulk pressure, $\pi^{\mu \nu}$ is the shear-stress tensor which is symmetric, 
traceless and orthogonal to $u^\mu$,  $V^\mu (=0, \text{here})$ is the baryon diffusion 4-current and 
$\Delta^{\mu \nu} = g^{\mu \nu}-u^\mu\, u^\nu$.  
The viscous terms obey the following evolution equations,
\begin{align}
u^{\alpha}\partial _{\alpha}\Pi &=-\frac{\Pi -\Pi _{NS}}{\tau _{\Pi}}-\frac{4}{3}\Pi\partial_{\alpha}u^{\alpha} ,\label{eqn:IS_bulk_relax}\\
\langle u^{\alpha}\partial _{\alpha}\pi ^{\mu \nu}\rangle &=-\frac{\pi ^{\mu \nu}-
\pi ^{\mu \nu}_{NS}}{\tau _{\pi}}-\frac{4}{3}\pi ^{\mu \nu}\partial_{\alpha}u^{\alpha},\label{eqn:IS_shear_relax}
\end{align}
where $\langle \cdot \rangle$ is defined as,
$$\langle A^{\mu \nu} \rangle=\left( \frac{1}{2}\Delta ^{\mu}_{\alpha}\Delta ^{\nu}_{\beta}+\frac{1}{2}\Delta ^{\nu}_{\alpha}\Delta ^{\mu}_{\beta}-\frac{1}{3}\Delta ^{\mu \nu}\Delta _{\alpha \beta}\right)A^{\alpha \beta},$$
$\Pi_{NS}$ and $\pi_{NS} ^{\mu \nu}$ are the Navier-Stokes limit of $\Pi$ and $ \pi^{\mu \nu}$ respectively, given by
\begin{equation}
\Pi_{NS} = -\zeta \theta \quad , \quad  \pi_{NS} ^{\mu \nu} = 2\eta \left\langle\partial ^\alpha u^\beta \right\rangle .
\end{equation}
The coefficients of shear ($\eta$) and bulk ($\zeta$) viscosities are positive, \emph{i.e.} $\eta ,\zeta > 0$. 

The hydrodynamical 
equations are solved in $(\tau,x,y,\eta_s)$ coordinates where,
$\tau=\sqrt{t^2-z^2}$ and $\eta_s=\text{tanh}^{-1}(z/t)$. 
The space-time evolution begins at time $\tau_0$. 
For lower energies the initial time, $\tau_0$ is taken as the time 
required by the nuclei 
to pass through one other ($\sim \frac{2R}{\gamma_zv_z}$) 
and for higher energies  ($\sqrt{s_{NN}}\geq 62.4$ GeV),
$\tau_0=1$ fm as shown in Table ~\ref{tab:tau0}. 
The initial energy density profile at $\tau_0$ is taken as:
\begin{equation}
\varepsilon (x,y,\eta_s;\tau_0) = e (x,y) \, f(\eta_s).
\end{equation}
A symmetric rapidity profile, $f(\eta_s)$, with the local energy-momentum conservation  puts a 
constraint on $e(x,y)$ as shown in Ref.~\cite{chunshen2020}. 
The energy deposited in the transverse plane, 
$e(x,y)$ depends on the 
number of wounded nucleons per unit area which 
has been calculated by using the optical Glauber model for
given impact parameter ($b$) at different $\sqrt{s_{NN}}$. 
The quantity, $e(x,y)$ is related to the 
the number of wounded nucleons per unit area in the transverse plane, $n_A(x,y)$ and $n_B(x,y)$ of
the colliding nuclei A and B respectively as:
$e(x,y)\propto 
\sqrt{n_A^2+n_B^2+2n_An_B\text{cosh}(2y_{\tt beam})}$,
where $y_{\tt beam}=\text{cosh}^{-1}(\sqrt{s_{NN}}/2)$. 
Here we consider Au+Au collisions
at $b=5.6$ fm for different $\sqrt{s_{NN}}$ that corresponds to 15-25\% centrality~\cite{supplement}.
The thickness function of the Au nucleus has been calculated by assuming Woods-Saxon profile for nuclear 
density with nuclear radius, $R_0=6.37$ fm, and surface thickness, $\delta=0.535$. The p+p inelastic cross-section, 
$\sigma^{\texttt{in}}_{NN}(\sqrt{s_{NN}})$, needed for the calculation of the number of wounded 
nucleons in the Glauber model has been taken from Refs.
~\cite{sigmaNN_parametrization1,sigmaNN_parametrization2}.

The initial velocity profile is taken as:
\begin{equation}
u^\mu (x,y,\eta_s; \tau_0)= \left( \cosh (\eta_s),0,0,\sinh (\eta_s)\right),
\end{equation}
The initial density profiles for energy and net baryon number have been 
computed with the parameters used in Ref.~\cite{chunshen2020}. 
The viscous terms have been initialized with their corresponding Navier-Stokes limit.
\begin{table}[H]
\caption{Values of $\tau_0$ used at different colliding energies}
\centering
\begin{tabular}{ |c |c |c |c |c |c |c |c |c |c| }
\hline
$\sqrt{s_{NN}}$ (GeV) & 14.5 & 19.6 & 27 & 39 & 62.4 & 200 \\
\hline
$\tau_0$ (fm) & 2.2 & 1.8 & 1.4 & 1.3 & 1.0 & 1.0 \\
\hline
\end{tabular}
\label{tab:tau0}
\end{table}
 
The EoS~\cite{parotto2020} employed here to solve the hydrodynamic equations 
reproduces the  lattice QCD results at zero baryon chemical potential.
The parameters $w$, $\rho$ and $\alpha_1$ that appear in the linear mapping from Ising
model to QCD in Ref.~\cite{parotto2020} have been fixed as $w=1$, $\rho=2$ and $\alpha_1=3.85^{\tt o}$.
The other parameters are same as Ref.~\cite{parotto2020}.
The transport coefficients are expected to 
diverge near the critical point following a scaling behavior~\cite{amonnai}:
$$\zeta \sim \xi^3 \quad , \quad \eta \sim \xi^{0.05}. $$
where $\xi (\mu,T)$  is the equilibrium correlation length, 
which is obtained through mapping QCD to 3D Ising model in the critical region. 
In the Ising model, $\xi$ is computed by taking the derivative of equilibrium magnetization, $M(r,h)$, 
with respect to the magnetic field, $h$, at fixed $r=(T-T_c)/T_c$, as~\cite{amonnai} 
$$\xi^2=\frac{1}{H_0}\left( \frac{\partial M}{\partial h}\right)_r,$$
where $H_0$ is a dimensionful parameter to get the correct dimensions of $\xi$. We shall take $H_0 = 1$ in our calculations
and $\left( \frac{\partial M}{\partial h}\right)_r$ is obtained from the EoS model~\cite{parotto2020} using chain rule of differentiation.
The extent of the critical domain
in the $\mu-T$ plane is determined by the condition: $\xi(\mu,T)=\xi_0$, where $\xi_0$ is taken
as 1.75 fm. 
The possibility of divergent behavior 
is incorporated through the following expressions of the transport coefficients~\cite{amonnai}
\begin{equation}
\label{eqn:crit_zeta}
\zeta = \zeta_0 \left( \frac{\xi}{\xi_0}\right)^3 \ , \ \eta = \eta_0 \left( \frac{\xi}{\xi_0}\right)^{0.05}
\end{equation}
Outside the critical region
the values  of the shear and bulk viscosities denoted  by 
$\eta_0,\ \zeta_0$  respectively   are  chosen as \cite{denicol2018,denicol2014}:
\footnotesize
\begin{align*}
\eta_0 (\mu,T) &= C_\eta \left( \frac{\varepsilon + p}{T}\right) \quad , \quad \zeta_0 (\mu,T) = 15 \ \eta_0 \left( \frac{1}{3}-c_s^2\right)^2
\end{align*}
\normalsize
The above parametrization is consistent with the estimates of the
temperature-dependent specific shear and bulk viscosity extracted using
Bayesian method~\cite{bernhard} away from the critical region. The authors of Ref.~\cite{Martinez}
compute the critical contribution to the bulk viscosity which is an order of
magnitude less than that of the noncritical contribution. 
The effect of reduced $\zeta_0$ in the critical region 
has been discussed in \cite{supplement}.

The dependence of the relaxation times appeared in Eqs.(\ref{eqn:IS_bulk_relax}) 
and (\ref{eqn:IS_shear_relax}) on $\xi$ are parameterized as:
\begin{equation}
\label{eqn:crit_relax}
\tau_\pi = \tau^0_\pi \left( \frac{\xi}{\xi_0}\right)^{0.05} \quad ,\quad \tau_\Pi = \tau^0_\Pi \left( \frac{\xi}{\xi_0}\right)^3.
\end{equation}
where $\tau^0_\pi$ and $\tau^0_\Pi$ are the relaxation times outside the critical region which are given 
by~\cite{denicol2018,denicol2014},
$$\frac{\tau^0_\pi}{5} =\tau^0_\Pi =\frac{C_\eta}{T}$$
with $C_\eta =0.08$.

%
\begin{figure}[H]
\centering
\includegraphics[scale=0.6]{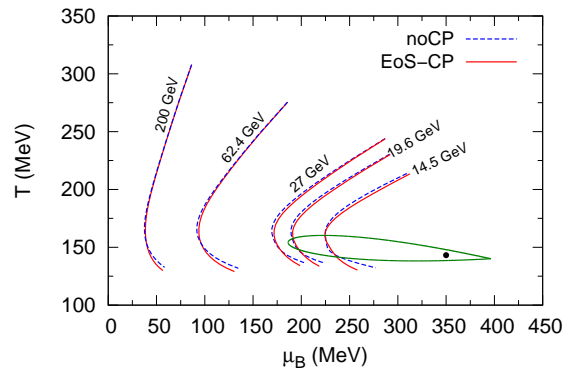}
\caption{Trajectories traced by the center of the fireball \emph{i.e.} 
$x=y=\eta_s=0$ in the $\mu-T$ plane for different $\sqrt{s_{NN}}$. 
The critical point is indicated by solid black dot at $(\mu,T)=(350,143.2)$ MeV. 
}
\label{fig1}
\end{figure}

The critical region and the trajectories traced by the center 
of the fireball in the ($\mu-T$) plane at different $\sqrt{s_{NN}}$ 
are shown in Fig.~\ref{fig1}. The black dot indicates the location
of the CP at $(\mu_c,T_c)=(350\text{MeV},143.2\text{MeV})$~\cite{parotto2020}.
The trajectories have been calculated by solving the hydrodynamic
equations with (denoted by EoS-CP) and without (noCP) the effects of CP. 
The CORNELIUS code is then used to find the freeze-out hyper surface $\Sigma_\varepsilon$, 
defined by 
$\varepsilon = 0.3 \text{GeV}/\text{fm}^3$. The spin polarization has been 
evaluated on this hyper surface.
The trajectories for $\sqrt{s_{NN}}=$14.5 GeV and 19.6 GeV pass through the critical
domain (shown by the closed contour in Fig.~\ref{fig1}) 
and those for higher
$\sqrt{s_{NN}}$ remain outside the critical domain. We evaluate the thermal vorticity
and subsequently the polarization of $\Lambda$ for system evolving along trajectories
passing through both inside and outside the critical domain. The effect of the CP on the 
polarization is expected to be larger for $\sqrt{s_{NN}}=14.5$ GeV as the trajectory
for this case is closer to the CP compared to other values of $\sqrt{s_{NN}}$ considered here.

\begin{figure}[H]
\centering
\begin{tabular}{cc}
\includegraphics[height=50mm,width=80mm,angle=0]{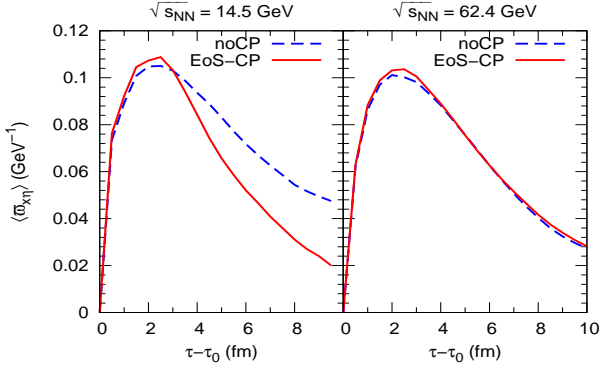} 
\end{tabular}
\caption{Time evolution of the mean of the  $x\eta$-component of thermal vorticity for 
$\sqrt{s_{NN}}=14.5 \text{GeV}$ and $62.4\text{GeV}$.}
\label{fig2}
\end{figure}

\begin{figure}[H]
\centering
\begin{tabular}{cc}
\includegraphics[height=60mm,width=70mm,angle=0]{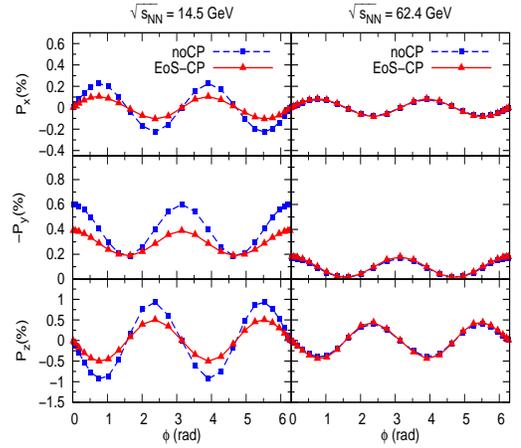} 
\end{tabular}
\caption{$x$, $y$ and $z$ components of $\Lambda$-polarization are plotted respectively in the upper, middle and the lower 
panels as a function of azimuthal angle in momentum space for 
$\sqrt{s_{NN}}=14.5 \text{GeV}\,\,\text{and}\,\,62.4 \text{GeV}$.}
\label{fig3}
\end{figure}

\begin{figure}[H]
\centering
\begin{tabular}{cc}
\includegraphics[scale=0.5]{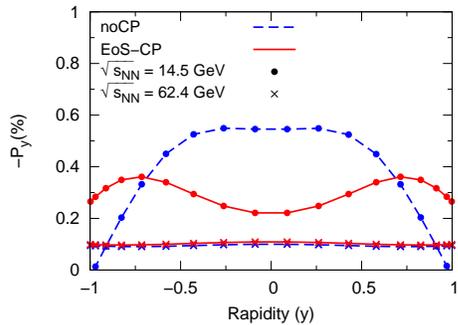} 
\end{tabular}
\caption{$y$-component of $\Lambda$-polarization plotted as a function of momentum rapidity 
for $\sqrt{s_{NN}}=14.5\,\,\text{GeV}\,\, \text{and}\,\,62.4\text{GeV}$.}
\label{fig4}
\end{figure}

\begin{figure}[H]
\centering
\includegraphics[scale=0.5]{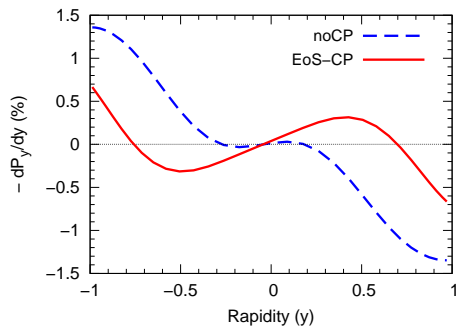}  
\caption{Negative slope of the $y$-component of the polarization of $\Lambda$-hyperon plotted as a function of rapidity.}
\label{fig5}
\end{figure}

\begin{figure}[H]
\centering
\includegraphics[scale=0.5]{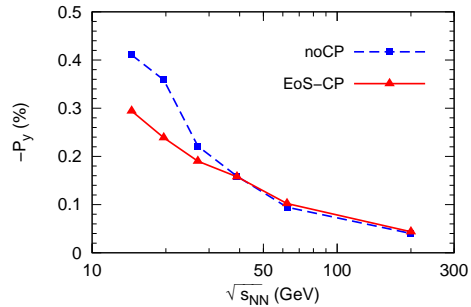}
\caption{Global polarization of $\Lambda$-hyperon plotted as a function of $\sqrt{s_{NN}}$.}
\label{fig6}
\end{figure}

The thermal vorticity at any space-time point of the fluid is given by~\cite{Becattini1,Becattini2}:
\begin{equation}
\varpi_{\mu\nu} = \frac{1}{2}\left[ \partial_\nu \beta_\mu - \partial_\mu \beta_\nu\right]
\end{equation}
where $\beta_\mu =u_\mu/T$. 
The time evolution of the $x\eta$ component of the thermal vorticity, averaged over the
spatial coordinates and weighted by the energy density, with and without the effects of CP
are shown in Fig.~\ref{fig2}. Initially the system has zero vorticity. The vorticity
generated by the viscous effects increase at first to attain some maximum value and
then decreases subsequently.
The evolution of the vorticity is affected by several factors. The 
hydrodynamic expansion does not create or destroy vortices but reduces it
through redistribution. The 
shear viscous coefficient is responsible for  its diffusion  and
the stretching and baroclinic torque enhance the vorticity.
Near the CP, absorption of sound wave affects the expansion directly and the diverging nature of 
the transport coefficients reduces the vorticity as seen in Fig.~\ref{fig2}. The observed
suppression of the vorticity due to CP at $\sqrt{s_{NN}}=14.5$ GeV is expected to
influence some of the experimental results. 
As the evolution trajectory for $\sqrt{s_{NN}}=62.4 \text{GeV}$ (and higher energies) 
remain outside the critical domain the results with and without the CP essentially overlap.  
The local thermal vorticity and  the $\Lambda$-polarization is calculated by using 
the following expression for mean spin vector of a spin-1/2 particle with 
four-momentum $p_\nu$~\cite{Smu} as,
$$S^\mu (x,p) = -\frac{1}{8m}(1-n_F)\epsilon^{\mu\nu\rho\sigma}p_\nu \varpi_{\rho \sigma} (x) + O (\varpi)^2$$
where $m$ is the mass of the particle, $\epsilon^{\mu\nu\rho\sigma}$ is the Levi-Civita tensor and $n_F$ is the Fermi-Dirac distribution. 
Since the mass of the $\Lambda$ is much larger than the 
temperature range being considered in this study,
we assume that $1-n_F\approx 1$ 
and $n_F\approx n_B$, where $n_B$ is the Boltzmann distribution. Consequently the expression for the mean spin vector becomes
$$S^\mu (x,p) = \frac{1}{8m}\epsilon^{\mu\nu\rho\sigma}p_\nu \partial_\rho \beta_\sigma $$
In the rest frame of the particle, the spin vector is $S^{*\mu} = (0,{\bf{S}^*})$, 
which is obtained by using the Lorentz transformation as:
$${\bf{S}^*} ={\bf{S}} - \frac{{\bf{p}}.{\bf{S}}}{E(E+m)}{\bf{p}}$$
The mean spin averaged over the surface $\Sigma$ is then  given by~\cite{Becattini1},
\begin{equation}
S^\mu (p) = \frac{\int d\Sigma_\lambda p^\lambda \ e^{-\beta (p.u-\mu)} S^\mu (x,p)}{\int d\Sigma_\lambda p^\lambda \ e^{-\beta (p.u-\mu)}}
\label{meanspin}
\end{equation}
The net spin is obtained by integrating over azimuthal angle ($0 \le \phi < 2\pi $),
rapidity ($|y| < 1$) and transverse momentum ($0<p_T<3 \text{ GeV}$) following the procedure of Ref.~\cite{wu2019}. 
Finally the spin polarization of $\Lambda$ is given by,
$${\bf{P}} = 2{\bf{S}^*}$$
In view of an ongoing puzzle on the issue of the variation of the longitudinal  
polarization with azimuthal angle ($\phi$) (~\cite{signpuzzle1,signpuzzle2,STARlambda}),
we display the variation of the
different components of the polarization, $P_x$, $P_y$ 
and $P_z$ with $\phi$ in Fig.~\ref{fig3}
for $\sqrt{s_{NN}}=14.5$ GeV and 62.4 GeV with
and without the effects of CP.
A systematic suppression of the polarization is observed ($xz$
is the reaction plane and $y$-axis is the axis of rotation here)
at $\sqrt{s_{NN}}=14.5$ GeV  which originates 
from several competing factors like enhancement of various transport coefficients, 
slower expansion, changes in baroclinic torque and vortex stretching  near the CP.
The polarization with and without the effect of CP overlap 
at $\sqrt{s_{NN}}=62.4$ GeV 
which is obvious as the trajectory for this case remains outside
the critical region (Fig.~\ref{fig1}). 
The variation of the $y$-component of the spin-polarization with rapidity
($y$) has been displayed in Fig.~\ref{fig4}. A drastic change is induced 
by the CP in the rapidity distribution of spin-polarization
at $\sqrt{s_{NN}}=14.5$ GeV. 
At $\sqrt{s_{NN}}=62.4$ GeV, 
the $P_y$ with and without CP is identical as expected. 
It is intriguing to note that the CP not only reduces  the polarization
around mid-rapidity  but also introduces 
strong qualitative changes in the slopes of the curves as shown in Fig.~\ref{fig5}.
Finally, on integration over  $p_T$, $\phi$ and $y$ 
the global polarization is obtained as a function of $\sqrt{s_{NN}}$.
The suppression of polarization is conspicuous   
for the trajectories passing through the critical domain at lower $\sqrt{s_{NN}}$ 
(Fig.~\ref{fig6}). The slope of the curve without CP 
is much steeper than the one with CP at lower $\sqrt{s_{NN}}$.
We have taken the initial orbital angular momentum (OAM) 
of the fireball as zero
resulting in smaller polarization compared to experimental value~\cite{LambdaStar}.
Inclusion of OAM  through the non-zero initial value of the
velocity profile~\cite{Becattini1} will enhance the magnitude of $-P_y$, however,
the difference in the polarization observed here with and without CP will still
persist. Results with the inclusion of OAM have been discussed
in the supplemental material~\cite{supplement}. 
The sensitivity of this result on other parameters has also been
presented in the supplemental material~\cite{supplement}.
\emph{Conclusions} - It is well-known that 
the  local vorticity of the fluid couples with the
quantum mechanical spin of the particles and polarize them. 
We have evaluated the spin polarization of $\Lambda$-hyperon with and without
the effects of CP and found a strong change in the spin polarization
around mid-rapidity as the system approaches the CP.
The thermal vorticity and consequently the polarization of the
$\Lambda$ hyperon for different colliding energies have been estimated and found
to be suppressed as the CP is approached. There are various physical 
processes which collectively contribute to the suppression. 
Although we have solved the relativistic equation to estimate
the vorticity, we consider below the evolution equation for kinematic 
vorticity ($\vec{\omega}=\nabla \times \vec{v}$)
for a compressible fluid with constant $\zeta$ and $\eta$
in the non-relativistic limit because in this form contributions
from various terms appear clearly, 
\begin{align*}
\frac{\partial \vec{\omega}}{\partial t} &= \left(\vec{\omega}\cdot \vec{\nabla}\right)\vec{v}- \left(\vec{v}\cdot \vec{\nabla}\right)\vec{\omega}-\theta\vec{\omega}
+\frac{1}{\rho^2}\vec{\nabla}\rho \times \vec{\nabla}p\\
&- \frac{1}{\rho^2}\left(\zeta +\frac{1}{3}\eta\right)\vec{\nabla}\rho \times \vec{\nabla}\theta
-\frac{\eta}{\rho^2}\vec{\nabla}\rho \times \nabla^2 \vec{v}+ \frac{\eta}{\rho} \nabla^2 \vec{\omega}.
\end{align*}
Here $\rho$ denotes the density of fluid and $\theta = \nabla \cdot \vec{v}$. As $\theta$ is a
measure of the expansion of the system, a larger expansion results in smaller vorticity as suggested
by the negative sign of the term $\theta\vec{\omega}$. The
terms depending on transport coefficients only, are written in the second line of the above equation.
The term proportional to $\nabla^2 \vec{\omega}$ is responsible for diffusion of vorticity in space,
the diffusion coefficient being $\frac{\eta}{\rho}$. The term of particular interest is proportional
to $\vec{\nabla}\rho \times \vec{\nabla}\theta$ which suggests that the vorticity dissipates if there is a
gradient in expansion rate for fluid cells i.e. the fluid cells having less density and expanding faster
will oppose the vorticity of the denser fluid cells expanding slowly. The strength of this effect
is proportional to $\left(\zeta +\frac{1}{3}\eta\right)$. 
It is clear that the suppression of vorticity and hence polarization, is a combined 
effect of the absorption of sound wave and the enhancement of various transport 
coefficients in presence of CP.
The drastic qualitative and quantitative changes induced
by CP in the rapidity distribution of $P_y$ can be used
to detect the CP experimentally
as the polarization of $\Lambda$ has already been measured 
by STAR collaboration~\cite{LambdaStar}. It is important to mention at this point
that the effects of CP on the $p_T$ spectra of the hadrons 
and on the $p_T$ and $y$ dependence of directed and elliptic flow
are found to be small~\cite{singh2022}. 

Some comments on the application of hydrodynamics near the CP are in
order here. Near the CP, the fluctuating modes do not relax faster than
the timescale of changes in slow/conserved variables due to which the local thermal equilibrium is not maintained 
making hydrodynamics inapplicable. The  validity of the hydrodynamics can, however,
be extended by adding a scalar variable representing the slow non-hydrodynamic modes
connected to  the relaxation  rate of the critical fluctuation
(see ~\cite{yin} and  \cite{stephanov2018} for details).
It has been explicitly shown that the modes associated with the scalar variable
lags behind the hydrodynamic modes resulting in back reactions on the
hydrodynamic variables~\cite{rajagopal}.
Further, it has been demonstrated in Ref.~\cite{rajagopal} that the  back reaction
has negligible effects on the hydrodynamic variables.
In view of this, the results presented in this work will be
useful in detecting the CP.
Moreover, we may also recall that if a system
is not too close to CP  then hydrodynamics can
still be applied in a domain around the CP~\cite{Stanley}.

We thank Sandeep Chatterjee and Tribhuban Parida for helpful discussions regarding the UrQMD transport code.

\begin{figure}[H]
\centering
\includegraphics[scale=0.45]{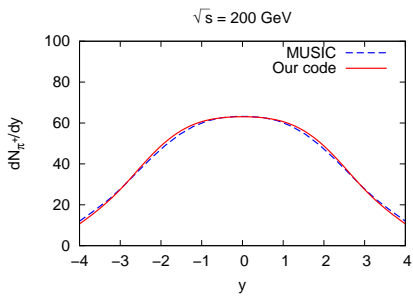}
\caption{The comparison of the rapidity distribution of $\pi^+$ from our code and the publicly available MUSIC code.}
\label{fig_music}
\end{figure}

\begin{figure}[H]
\centering
\includegraphics[scale=0.45]{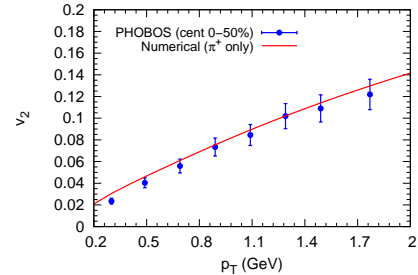}
\caption{Our numerical result on $p_T$ dependence of elliptic flow, $v_2$, compared to the experimental data from Ref.~\cite{phobos2005}.}
\label{fig_v2pT}
\end{figure}

\newpage
\noindent{\section{Supplementary Material}}
In this supplemental material, we present a few test results from our hydrodynamic code
with the inclusion of the effects of critical point (CP) through the equation of state (EoS) 
and scaling behaviour of transport coefficients, and
contrast the results without CP (denoted as noCP in the text and figures). 
We shall also discuss the effect of non-zero orbital angular momentum (OAM) on our results. 
We have already shown a comparison of our numerical results with the analytical Gubser solution in (2+1) 
dimensions in Ref.~\cite{singh2022}. There is no analytical result available in (3+1)-dimensions. 
Therefore, to test the code, we compare our result on the rapidity distribution of positively charged 
pion with the output of publicly available MUSIC code~\cite{musiccode} without the resonance decays 
in Fig.~\ref{fig_music}. For the next check, we reproduce the PHOBOS data on $p_T$ dependence of elliptic flow in 0-50\% centrality of Au+Au collisions at $\sqrt{s_{NN}}=200$ GeV~\cite{phobos2005} in Fig.~\ref{fig_v2pT}.

\begin{figure*}[t]
\centering
\includegraphics[scale=0.45]{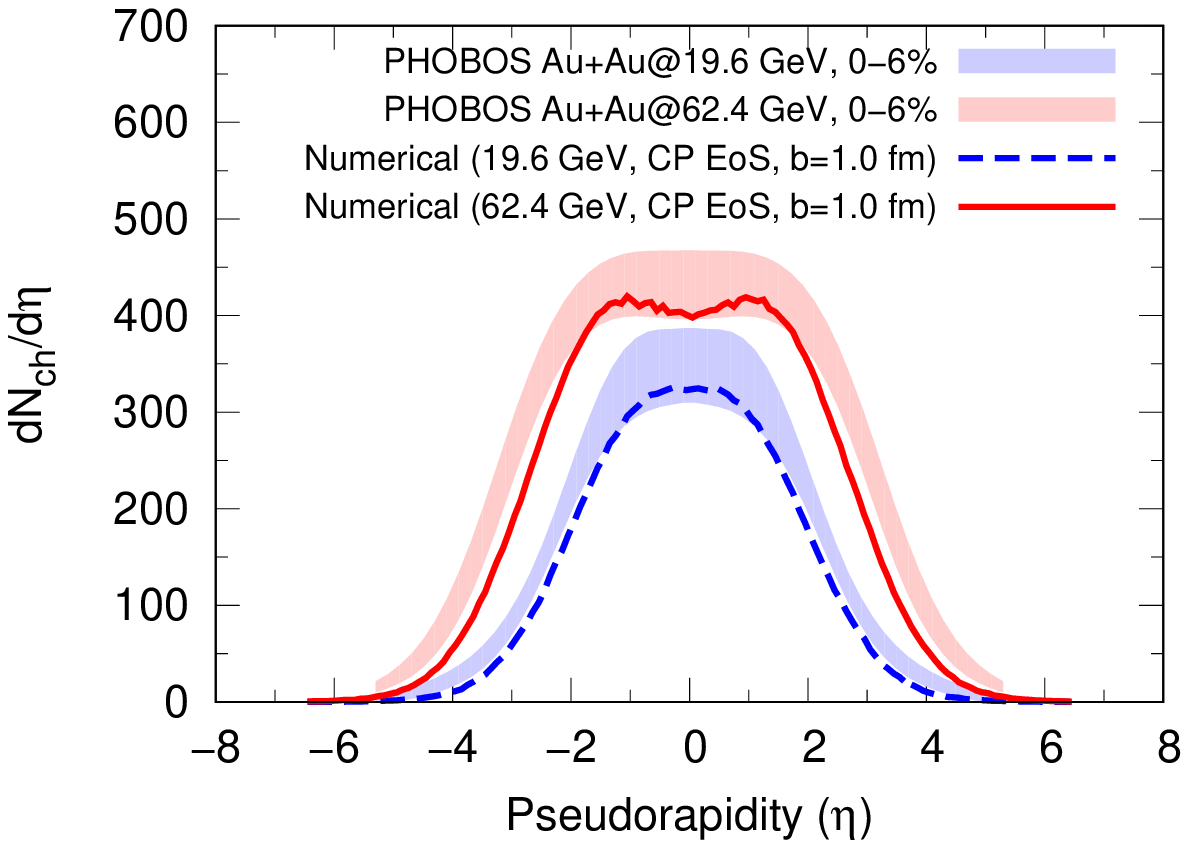}
\includegraphics[scale=0.45]{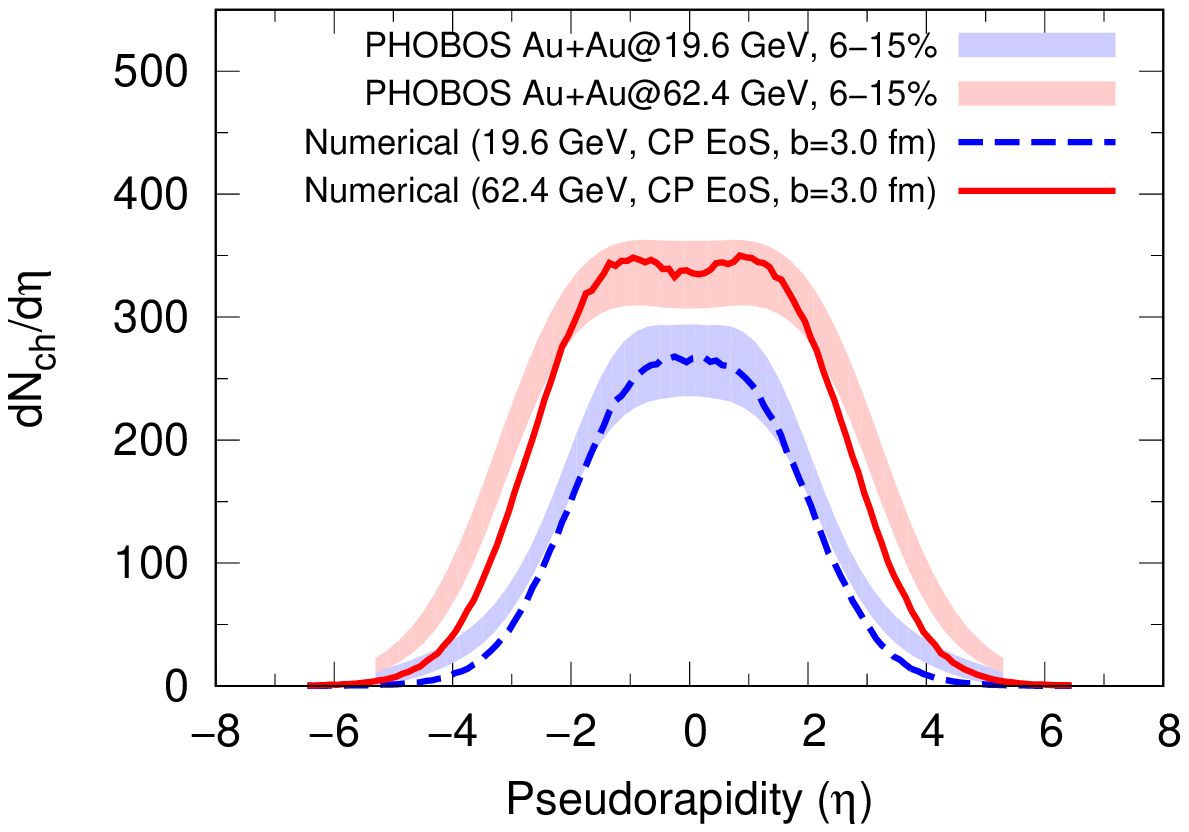}
\includegraphics[scale=0.45]{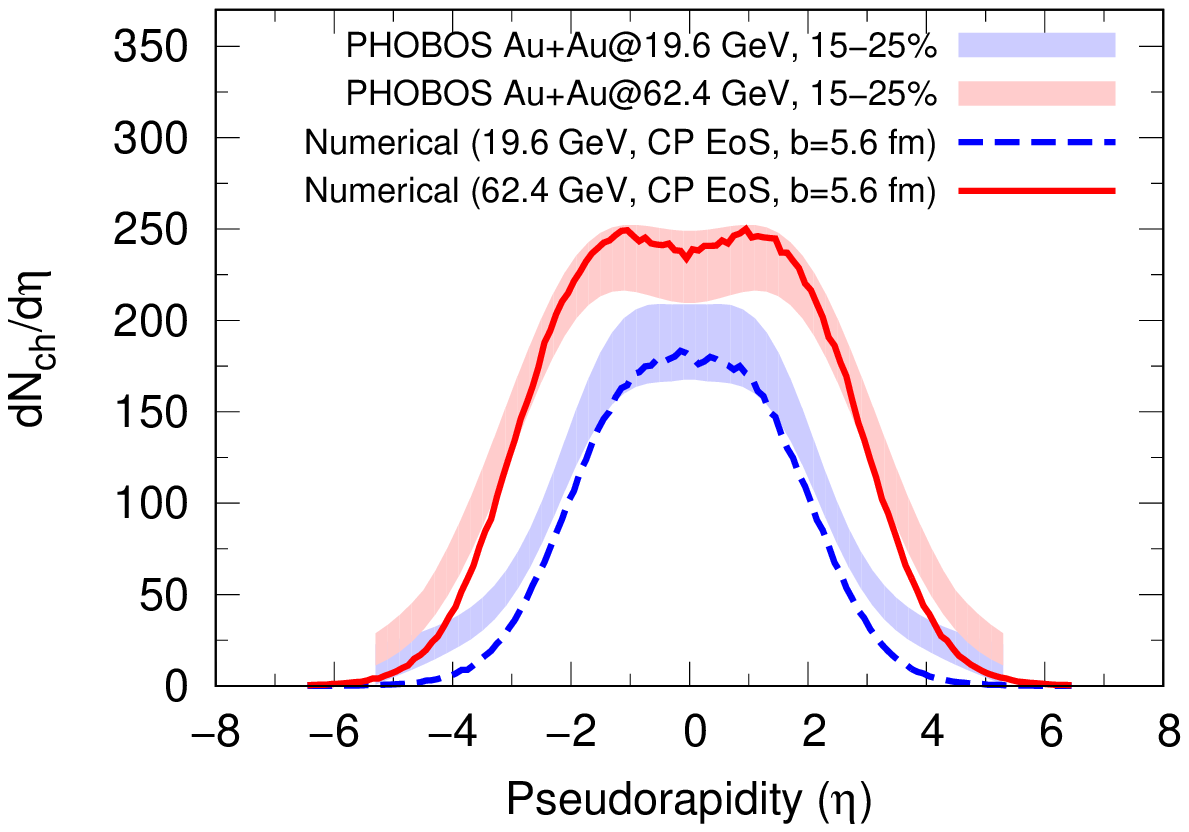}
\caption{The numerical result for $dN_{\text{ch}}/d\eta$ is compared with experimental data by PHOBOS in different centralities~\cite{phobos2003,phobos2006}.}
\label{fig_dnchdeta}
\end{figure*}

\begin{figure*}[t]
\includegraphics[scale=0.45]{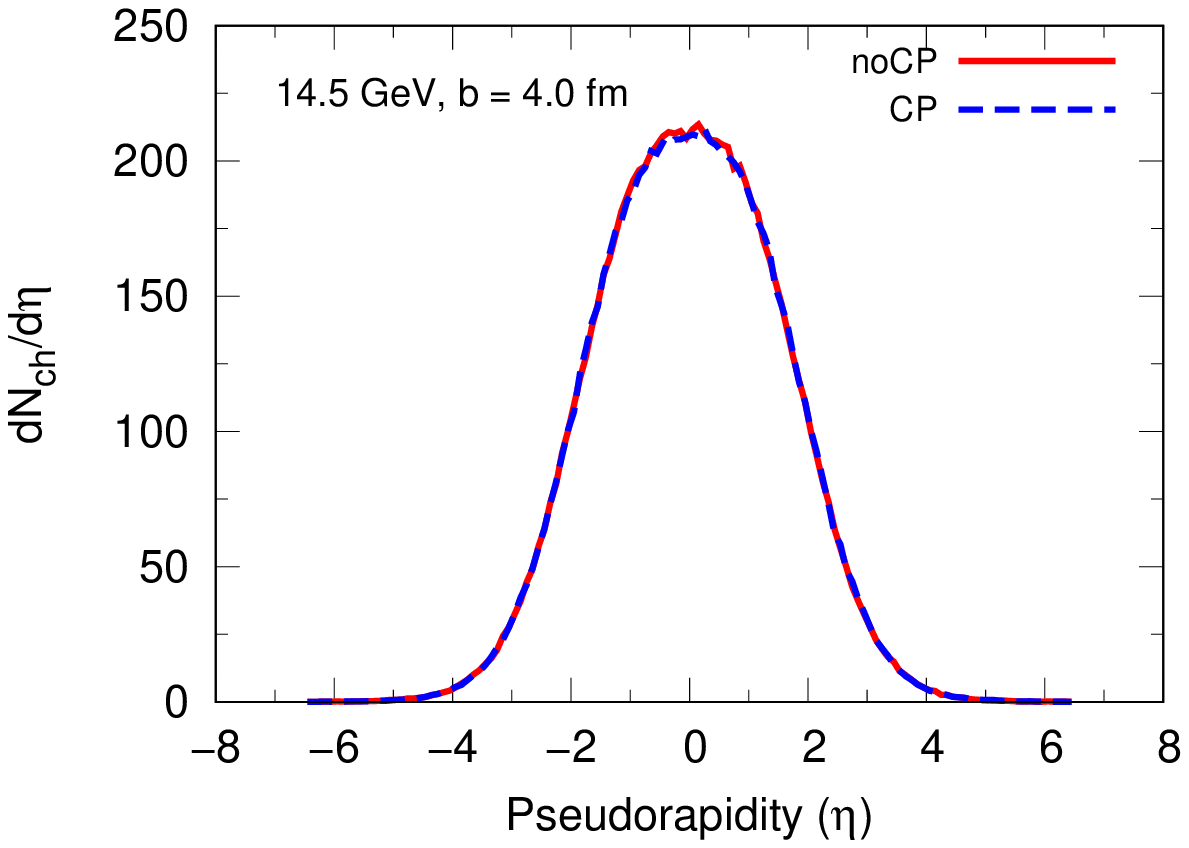}
\includegraphics[scale=0.45]{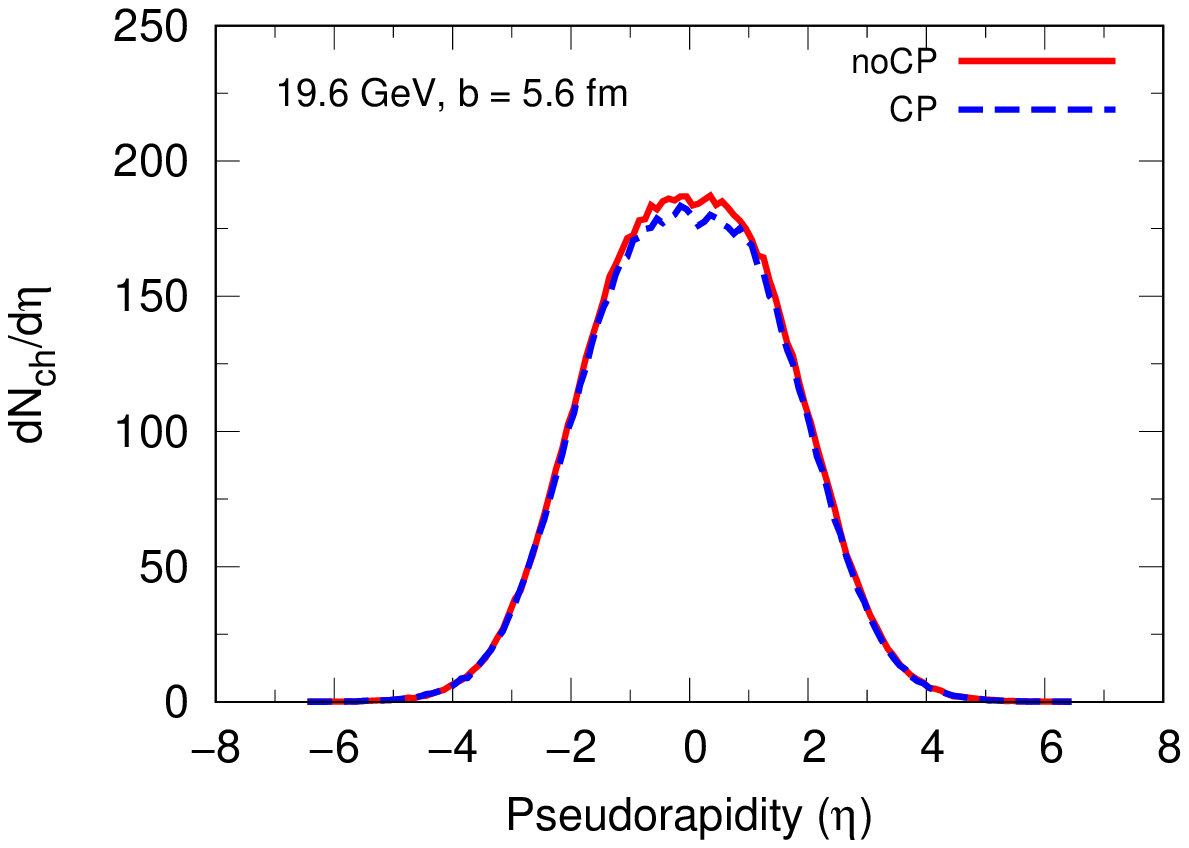}
\caption{$dN_{\text{ch}}/d\eta$ shown as a function of pseudorapidity with (CP) and without (noCP) the critical point in EoS for colliding energies (left) 14.5 GeV at impact parameter $b=4$ fm, and (right) 19.6 GeV at impact parameter $b=5.6$ fm.}
\label{fig_cpnocp}
\end{figure*}

We next reproduce the charged particle pseudorapidity distribution for two colliding 
energies ($\sqrt{s_{NN}}$) and different centralities in Fig.~\ref{fig_dnchdeta}. To generate 
the plots in Fig.~\ref{fig_dnchdeta}, we use the switching energy density 
$\varepsilon_{\text{sw}}=0.3$ GeV/fm$^3$. The constant energy density 
hypersurface ($\varepsilon_{\text{sw}}$) is obtained using the CORNELIUS code~\cite{cornelius} which is 
then given as input to the UrQMD transport code~\cite{urqmd} (which does not
include spin effects) and generate 1000 events. 
It should be mentioned here that the width of the experimental distribution is 
slightly underestimated because we have used a single impact parameter and not an 
event-by-event simulation that would consist of a mixture of several impact parameters. 
The results of Fig.~\ref{fig_dnchdeta} include the effects due to CP.
However, on comparing with the results without CP, the effect is negligible
as demonstrated in Fig.~\ref{fig_cpnocp} (see also Ref.~\cite{singh2022}).

Now we implement a non-zero OAM in the initial condition. The initial condition model that we use from Ref.~\cite{shen2020} has been generalized to include a non-zero OAM in Ref.~\cite{shen2021}. This is done by introducing a parameter, $f$, that takes value in the interval [0,1] and it controls the fraction of longitudinal momentum that can be attributed to the flow velocity. $f=0$ corresponds to the Bjorken flow scenario. The assumption for the initial energy-momentum current in Ref.~\cite{shen2021} can be achieved through the following choice of rest frame quantities: 
\begin{align*}
p&=\varepsilon \ , \ u^x=u^y=0,\\
u^\tau &= \cosh\left( \frac{y_L}{2}\right)\ , \  u^\eta =\frac{1}{\tau_0}\sinh\left( \frac{y_L}{2}\right),
\end{align*}
where $p$, $\varepsilon$, and $u^\mu$, respectively, denote the pressure, the energy density, and the fluid four flow-velocity. Also, $y_L=fy_{\text{CM}}$ denotes the local longitudinal rapidity variable and $y_{\text{CM}}$ is the local center-of-mass rapidity variable (defined in the main article). 
The components of the initial energy-momentum tensor then take the following form:
\begin{align*}
T^{\tau\tau} &= (\varepsilon+p)(u^\tau)^2 - p = \varepsilon \cosh(y_L),\\
T^{\tau\eta} &= (\varepsilon+p)u^\tau u^\eta  = \frac{\varepsilon}{\tau_0} \sinh(y_L),\\
T^{\eta\eta} &= (\varepsilon+p)u^\eta u^\eta  = \frac{\varepsilon}{\tau_0^2} \cosh(y_L),\\
\end{align*}
which is consistent with the assumption of Ref.~\cite{shen2021}.
The trace of the energy-momentum tensor is given by
$$T^\mu_\mu = T^{\tau \tau} - \tau^2T^{\eta\eta} = 0.$$
Hence, our choice for the rest frame quantities does not violate the condition for the trace 
of the energy-momentum tensor. 
The viscous stresses are initialized to their corresponding Navier-Stokes limit,
the expressions of which are:
\begin{align*}
\pi^{\tau\tau}&=-2\eta \tau u^\tau (u^\eta)^2 - \frac{2\eta}{3\tau} u^\tau \left[ 1- (u^\tau)^2\right] ,\ \pi^{\tau x} = -\eta\partial_{x}u^\tau ,\\
\pi^{\tau y} &= -\eta\partial_{y}u^\tau,\ \pi^{\tau \eta} =-\eta \tau (u^\eta)^3 -\frac{\eta u^\eta}{\tau}-\frac{\eta}{3\tau}(u^\tau)^2u^\eta,\\
\pi^{xx } &= \frac{2\eta}{3\tau }u^\tau,\ \pi^{xy }=0 ,\ \pi^{x\eta }= -\eta \partial_x u^\eta, \ \pi^{yy } = \frac{2\eta}{3\tau}u^\tau,\\
\pi^{y\eta } &= -\eta \partial_{y}u^\eta,\ \pi^{\eta\eta } = -\frac{4\eta}{3\tau}u^\tau \left[ \frac{1}{\tau^2}+(u^\eta)^2\right]
\end{align*}

It was stated in Ref.~\cite{shen2021} that the parameter $f$ 
has negligible effects on most of the global observables such as the pseudorapidity distributions, 
particle yields, and elliptic flow. We have checked this and conclude the same. The rest of the 
analysis that follows is carried out on a constant energy density hypersurface, 
$\varepsilon = 0.3$ GeV/fm$^3$, 
we shall denote this surface as $\Sigma _\varepsilon$ below.

\begin{figure}[H]
\centering
\includegraphics[scale=0.45]{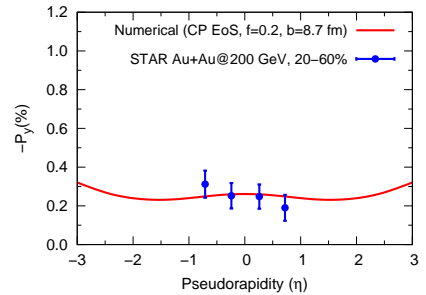}
\caption{The numerical result for $y$-component of $\Lambda$-polarization as function of pseudorapidity is compared with the experimental data by STAR from Au+Au collisions at $\sqrt{s_{NN}}=200$ GeV in 20\%-60\% centrality~\cite{star2018}. Vertical lines are the statistical uncertainties only.}
\label{fig_phvseta}
\end{figure}

\begin{figure}[H]
\centering
\includegraphics[scale=0.45]{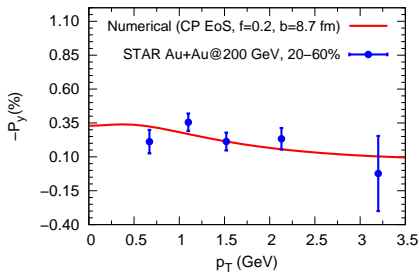}
\caption{The numerical result for $y$-component of $\Lambda$-polarization as function of transverse momentum ($p_T$) is compared with the experimental data by STAR from Au+Au collisions at $\sqrt{s_{NN}}=200$ GeV for 20-60\% centrality~\cite{star2018}. Vertical lines are the statistical uncertainties only.}
\label{fig_phvspt}
\end{figure}

\begin{figure}[H]
\centering
\includegraphics[scale=0.45]{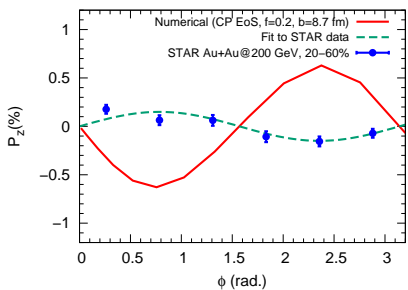}
\caption{Azimuthal angle dependence of $z$-component of spin polarization of $\Lambda$-hyperon. STAR measurements are from Ref.\cite{star2019}.}
\label{fig_phvsphi}
\end{figure}

By setting $f=0.2$, impact parameter, $b=8.7$ fm, and using the same set of values for other parameters of the IC model for Au+Au collision at $\sqrt{s_{NN}}=200$ GeV, described in the main article, we compute the negative $y$-component of global polarization, $-P_y$, of $\Lambda$-hyperon for 20-60\% centrality. Our result is $-P_y$=0.254\%. The corresponding experimental value from STAR is 0.277$\pm$ 0.040 (stat) and using the updated PDG value of 
 $\alpha_\Lambda$ is 0.243$\pm$ 0.035\% (stat)~\cite{star2018}. We also show the pseudorapditiy and transverse momentum dependence in the same centrality in Figs.\ref{fig_phvseta} and \ref{fig_phvspt}. Again, we use EoS with 
CP to generate these results. Because the difference between CP and noCP equation of states is negligible at such large colliding energy. The azimuthal angle dependence of the longitudinal component of the spin-polarization, $P_z$, is shown in Fig.~\ref{fig_phvsphi}. The sign of our numerical results is opposite to that of experimental data. This problem is known as the longitudinal sign puzzle in the literature (see~\cite{becattini2020} for a review).

Having validated our code we now carry out our simulation near the critical point at colliding 
energy $\sqrt{s_{NN}}=14.5$ GeV.  
The negative $y$-component of the global polarization is suppressed in the presence of CP 
as compared to the case when the CP is absent for a given value of $f$. This illustrates 
that the values of global polarization with and without CP are different at fixed $f$ which
is clearly observed in the main article (Fig. 5) for zero initial OAM (corresponding to $f=0$). 
We carry out an exercise to  verify whether the same global polarization of 
$\Lambda$ hyperon can be obtained 
with and without CP by tuning the parameter $f$ in the presence of OAM.
The values $f=0.45$ with  CP and $f=0.53$ without CP corresponds to the same
value of $-P_y$=0.92\% of $\Lambda$-hyperon for $b=5.6$ fm which nearly reproduces 
the experimental value measured by the STAR collaboraion for 20\%-60\% centrality.
This indicates that the data on global polarization can not be used to exclusively 
determine the CP effects, because other parameters can be tuned to the data.

\begin{figure}[H]
\centering
\includegraphics[scale=0.42]{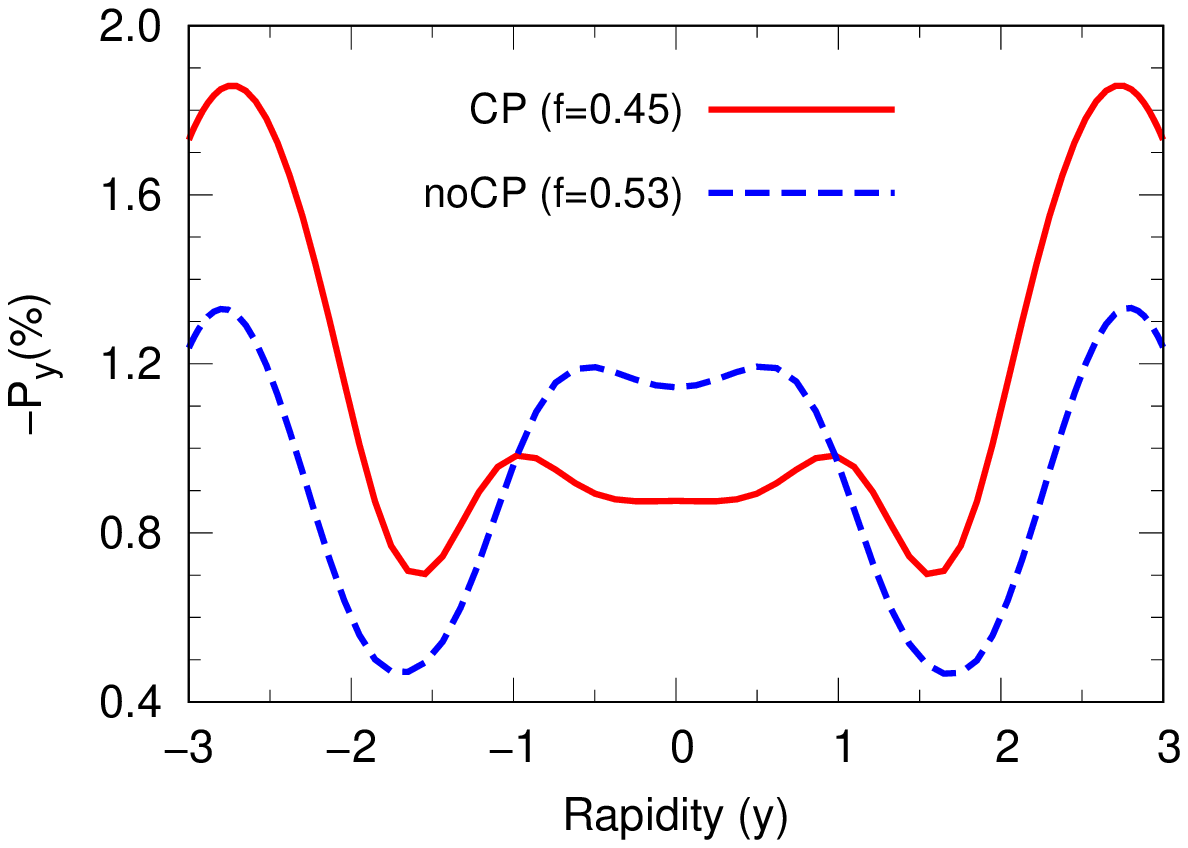}\\
\includegraphics[scale=0.45]{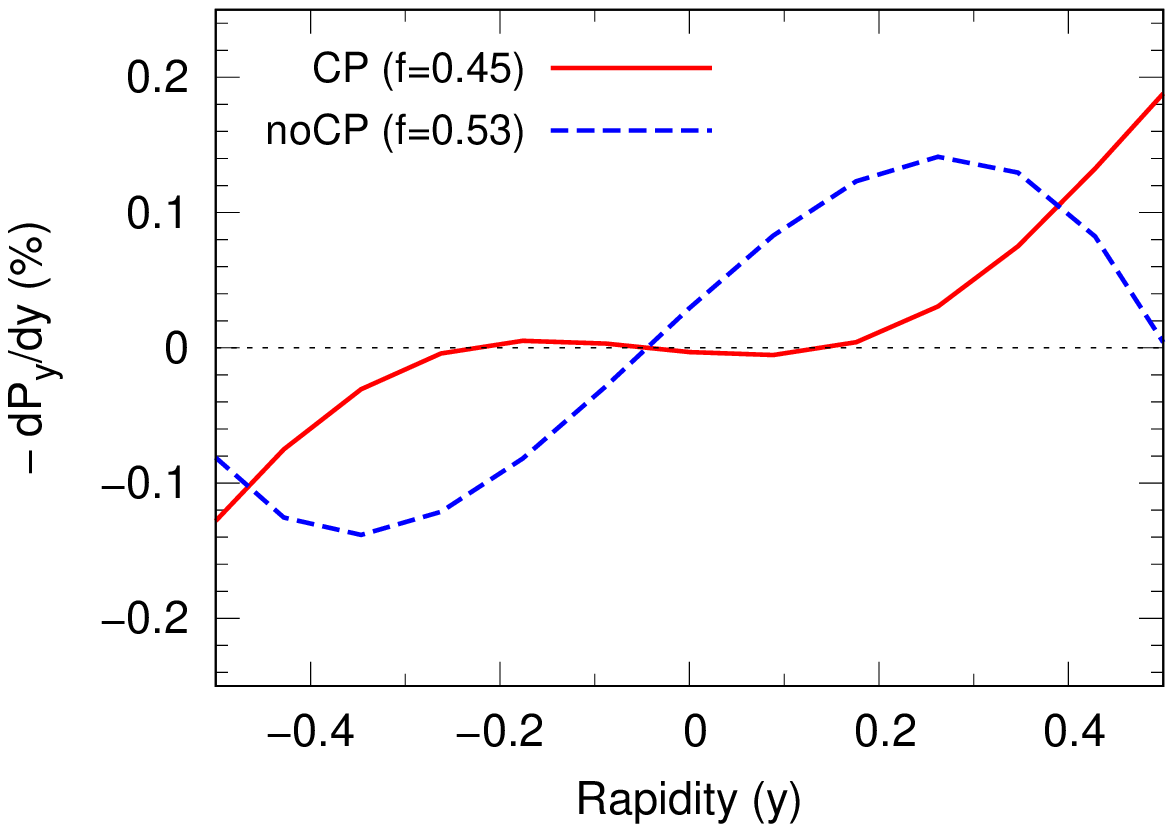}
\caption{The top panel shows rapidity dependence of $y$ component of $\Lambda$-hyperon polarization at $\sqrt{s_{NN}}=14.5$ GeV and $b=5.6$ fm with (CP) and without (noCP) critical point EoS, and the bottom panel shows the derivative as a function of rapidity.}
\label{fig_phvsy}
\end{figure}

However, once the value of $f$ is tuned to the global polarization data then the 
use of the same value of $f$ (which fixes the OAM) predicts significant difference in the rapidity
distribution of $P_y$ with and without CP, clearly indicating that 
the rapidity distribution of $-P_y$ is sensitive to CP. 
We plot the rapidity distribution of $-P_y$ in the top panel of Fig.~\ref{fig_phvsy}. 
We observe a suppression of about $25\%$ in $-P_y$ at mid-rapidity. The change in other 
observables like $p_T$ spectra, elliptic flow, $dN_{\text{ch}}/dy$ is at most 8\% on the 
surface $\Sigma_\varepsilon$~\cite{singh2022}. We also compute the derivative of $P_y$ with 
respect to rapidity and plot as a function of rapidity in the bottom panel of Fig.~\ref{fig_phvsy}. 
We observe a slight negative slope for $-dP_y/dy$ at mid-rapidity opposite to the case when there 
is no critical point. We cannot confirm the negative sign of the slope by further 
approaching the critical point due to the limitations of the EoS model which is valid for $\mu_B<450$ MeV.
Beyond 450 MeV, the speed of sound starts to give unphysical results. So close to CP, many fluid cells whose
trajectories cross $\mu_B=450$ MeV become problematic.
However, it should be mentioned that the suppression that we observe is when the center of the fireball
created in $\sqrt{s_{NN}}=14.5$ GeV energy is still 100 MeV away from the critical point along the $\mu_B$ axis. 
We expect the effect to get enhanced on further approach toward the critical point. 

The sensitivity of the spin polarization to the EoS can be 
understood from the following expression for the spin polarization in the rest frame of $\Lambda$-hyperon at any point on $\Sigma_\varepsilon$~\cite{becattini2020}:
$$\vec{S}^*(x,p) \propto \frac{\gamma}{T^2}\vec{v}\times \nabla T+\frac{1}{T}\left( \vec{\omega}-(\vec{\omega}\cdot \vec{v})\vec{v}\right)+\frac{1}{T}\gamma \vec{A}\times \vec{v},$$
where $\vec{\omega}$ is the vorticity ($=\nabla \times \vec{v}$), $\gamma$ is the Lorentz factor
 and $A$ denotes the acceleration of the fluid element. In nutshell, spin polarization depend on 
the gradients of temperature and curl of flow-velocity $(=\omega$). These gradients in turn depend 
on the expansion dynamics of the system and the expansion is strongly influenced 
by the speed of sound which 
is obtained from EoS. Since the sound wave gets suppressed at the critical point, the system 
undergoes a slow expansion that results into smaller gradients of temperature and flow-velocity, 
as they will not change much. 
This should then result into a suppression of spin-polarization as confirmed by our 
simulations.  
The effects of CP on other observables have been presented in Ref.~\cite{singh2022} by 
solving the same hydrodynamic equations with and without the critical point.

\begin{figure}[t]
\centering
\includegraphics[scale=0.42]{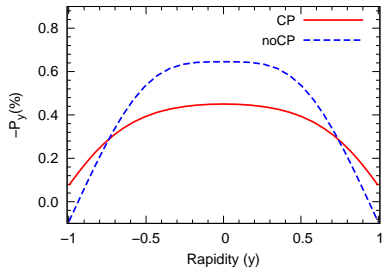}
\caption{Suppression in $y$-component of spin polarization for $f=0$, $\zeta'_0 =\zeta _0/10$ and $\xi'_0=0.8$ fm.}
\label{fig_zetaby10}
\end{figure}

To further demonstrate the robustness of our prediction, we show in Fig.~\ref{fig_zetaby10} the suppression in polarization when the bulk viscosity away from critical region, denoted by $\zeta_0$ in Eq.(9) of main article,
is decreased by a factor of 10 i.e. $\zeta'_0=\frac{\zeta_0}{10}$ and the length scale, $\xi_0$
in Eq.(9) of main article, that marks the boundary of the critical region is taken as $\xi'_0=0.8$ fm. We still see a suppression of about 30\% at mid-rapidity.


\begin{thebibliography}{99}

\bibitem{bazavov} A. Bazavov, F. Karsch, S. Mukherjee and 
P. Petreczky (USQCD Collaboration), arXiv:1904.09951 [hep-lat].







\bibitem{Fodor:2004nz} Z.~Fodor and S.~Katz, JHEP \textbf{04}, 050 (2004).

\bibitem{lqcd} H. T. Ding, F. Karsch and S. Mukherjee, Int. J. Mod.
Phys. E {\bf 24}, 1530007 (2015).

\bibitem{stephanov} M. A. Stephanov, K. Rajagopal and E. V. Shuryak,
Phys.  Rev. Lett. {\bf 81}, 4816 (1998).

\bibitem{nxu} X. Luo and N. Xu, Nucl. Sci. Tech. {\bf 28}, 112 (2017).

\bibitem{yyin1} J. Brewer, S. Mukherjee, K. Rajagopal and Yi Yin, Phys. Rev. C {\bf 98}, 061901 (2018).

\bibitem{stephanovprl1} M. A. Stephanov,  Phys. Rev. Lett.,  {\bf 102}, 032301 (2009). 

\bibitem{stephanovprl2} M. A. Stephanov,  Phys. Rev. Lett.,  {\bf 107}, 052301 (2011). 

\bibitem{yyin} Y. Yin, arXiv:1811.06519 [nucl-th].

\bibitem{LambdaStar} L. Adamczyk {\it et al.} (for STAR collaboration),
Nature {\bf 548}, 63 (2017).  

\bibitem{becattini2020} F. Becattini and M. A. Lisa, Ann. Rev. Nucl. Part. Sci. 70 (2020) 395.

\bibitem{barnett} S. J. Barnett, Phys. Rev. {\bf 6}, 239 (19150.

\bibitem{deHaas} A. Einstein and W. J. de-Haas, Ver. Dtsch. Ges. {\bf 17},
152 (1915).

\bibitem{ztliang} Z. T. Liang and X. N. Wang, Phys. Rev. Lett. {\bf 94}, 102301 (2005); 
Z. T. Liang and X. N. Wang, Phys. Rev. Lett. {\bf 96}, 039901 (2005). 

\bibitem{spinhydro} R. Takahashi {\it et al.}, Nat. Phys. {\bf 12},
52 (2016).

\bibitem{supplement} S. K. Singh and J. Alam, Supplemental material. 

\bibitem{hasan} Md Hasanujjaman, M. Rahaman, A. Bhattacharyya and
J. Alam, Phys. Rev. C {\bf 102}, 034910 (2020).

\bibitem{cve} D. E. Kharzeev, J. Liao, S. A. Voloshin and G. Wang,
Prog. Part. Nucl. Phys. {\bf 88}, 1 (2016). 

\bibitem{karpenko2014} Iu. Karpenko \emph{et al.}, 
Comput. Phys. Commun. 185 (2014) 3016–3027.

\bibitem{chunshen2020} C. Shen and S. Alzhrani, Phys. Rev. C {\bf 102}, 014909 (2020).

\bibitem{parotto2020} P. Parotto \emph{et al.}, Phys. Rev. C 101, 034901 (2020).

\bibitem{Cornelius} P. Huovinen and H. Petersen, Eur. Phys. J. A 48, 171 (2012).

\bibitem{gubser2010} S. S. Gubser, Phys. Rev. D \textbf{82}, 085027 (2010).

\bibitem{azhydro} P. F. Kolb, J. Sollfrank and U. Heinz, Phys. Rev. C {\bf 62}, 054909 (2000).

\bibitem{music} B. Schenke, S. Jeon, C. Gale, Phys. Rev. C {\bf 82}, 014903 (2010).

\bibitem{sigmaNN_parametrization1}  J. Cudell \emph{et al.} (COMPETE), Phys. Rev. Lett. {\bf 89}, 201801 (2002).

\bibitem{sigmaNN_parametrization2} B. Abelev \emph{et al.} (ALICE Collaboration) Phys. Rev. C {\bf 88}, 044909 (2013).

\bibitem{amonnai} A. Monnai, S. Mukherjee and Y. Yin, Phys. Rev. C {\bf 95}, 034902 (2017).

\bibitem{denicol2018} G. S. Denicol, C.  Gale, S. Jeon, A.  Monnai, B.  Schenke, and C. Shen, 
Phys. Rev. C {\bf 98}, 034916 (2018).

\bibitem{denicol2014} G. S. Denicol, S. Jeon and C. Gale, Phys. Rev. C {\bf 90}, 024912 (2014).

\bibitem{bernhard} J. E. Bernhard, J. S. Moreland and S. A. Bass, Nat. Phys. 15, 1113-1117 (2019)

\bibitem{Martinez} M. Martinez, T. Sch\"afer and V. Skokov, Phys. Rev. D {\bf 100}, 074017 (2019). 

\bibitem{Becattini1} F. Becattini,  {\it et al.}, Eur. Phys. J. C {\bf 75}, 406 (2015).

\bibitem{Becattini2} F. Becattini, Iu. Karpenko, M. A. Lisa, I. Upsal and S. A. Voloshin, 
Phys. Rev. C {\bf 95}, 054902 (2013); F. Becattini, L. P. Csernai and D. J. Wang, 
Phys. Rev. C {\bf 88}, 034905 (2013).

\bibitem{Smu} F. Becattini, V. Chandra, L. D. Zanna and E. Grossi, Ann. Phys. {\bf 338}, 32 (2013);
R.-h. Feng, L.-g. Pang, Q. Wang and X. N. Wang, Phys. Rev. C {\bf 94}, 024904 (2016).  

\bibitem{wu2019} H. Z. Wu, L. G. Pang, X. G. Huang and Q. Wang, Phys. Rev. Research. 1, 033058 (2019).

\bibitem{signpuzzle1} F. Becattini and Iu. Karpenko, Phys. Rev. Lett. 120, 012302 (2018).

\bibitem{signpuzzle2} Iu. Karpenko, Lecture Notes in Physics, vol. 987, Springer (2021) 247-280 [arXiv:2101.04963].

\bibitem{STARlambda} J. Adam \emph{et al.} (STAR Collaboration), Phys. Rev. Lett. 123, 132301 (2019).
\bibitem{singh2022} S. K. Singh  and J. Alam,  arXiv:2205.14469 [nucl-th]

\bibitem{yin} M. Stephanov and Y. Yin, Nucl. Phys. A {\bf 967},  876 (2017).

\bibitem{stephanov2018} M. Stephanov and Y. Yin, Phys. Rev. D \textbf{98}, 036006 (2018).

\bibitem{rajagopal} K. Rajagopal,  G. W. Ridgway,  R. Weller and Y Yin, Phys. Rev. D {\bf 102}, 094025 (2020).

\bibitem{Stanley} H. E. Stanley, Introduction to phase transitions and critical phenomena, Oxford University Press, 1971.


\end{thebibliography}

\begin{thebibliography}{99}
  \bibitem{singh2022} S. K. Singh and J. Alam, arXiv:2205.14469.
  \bibitem{musiccode} http://www.physics.mcgill.ca/music/
  \bibitem{phobos2005} B. B. Back \emph{et al.}, Phys. Rev. C 72 (2005) 051901.
  \bibitem{cornelius} P. Huovinen and H. Petersen, Eur. Phys. J. A 48, 171 (2012).
  \bibitem{urqmd} H. Petersen, J. Steinheimer, G. Burau, M. Bleicher and H. Stöcker, Phys. Rev. C 78 (2008) 044901.
  \bibitem{shen2020} C. Shen and S. Alzhrani, Phys. Rev. C 102, 014909 (2020).
  \bibitem{shen2021} S. Ryu, V. Jupic, and C. Shen, Phys. Rev. C 104, 054908 (2021).
  \bibitem{star2018} J. Adam \emph{et al.} (STAR Collaboration), Phys. Rev. C 98, 014910 (2018).
\bibitem{becattini2020} F. Becattini and M. A. Lisa, Ann. Rev. Nucl. Part. Sci. 70 (2020) 395.
  \bibitem{phobos2003} B.B.Back \emph{et al.} (PHOBOS Collaboration), Phys. Rev. Lett.91, 052303 (2003).
  \bibitem{phobos2006} B.B.Back \emph{et al.} (PHOBOS Collaboration), Phys. Rev. C 74, 021901(R) (2006).
  \bibitem{star2019} J. Adam \emph{et al.} (STAR Collaboration), Phys. Rev. Lett. 123, 132301 (2019).
\end{thebibliography}
\end{document}